\documentclass{article}

\usepackage{PRIMEarxiv}

\usepackage[utf8]{inputenc} 
\usepackage[T1]{fontenc}    
\usepackage{hyperref}       
\usepackage{url}            
\usepackage{booktabs}       
\usepackage{amsfonts}       
\usepackage{nicefrac}       
\usepackage{microtype}      
\usepackage{lipsum}
\usepackage{fancyhdr}       
\usepackage{graphicx}       
\graphicspath{{media/}}     

\title{QVNTVS, Open-Source Quantum Well Simulator

}

\author{
  Barbaros Şair \\
  Undergraduate Researcher at UNAM, National Nanotechnology Research Centre \\
  Senior Student in Electrical and Electronics Engineering, Middle East Technical University \\
  Ankara, Turkey\\
  sair.barbaros@metu.edu.tr\\
  QVNTVS Source Code : \url{https://github.com/sairbarbaros/QVNTVS}
}

\begin{document}
\maketitle

\begin{abstract}
Quantum Wells (QW) are of great importance in optoelectronic devices such as LEDs and LASERs, being the emissive layers. Simulating the quantum particles in different QW topologies like rectangular finite potential wells, multiple potential wells, and triangular biased potential well heterojunctions enables faster modeling, theoretical characterization, and more. QVNTVS performs energy level and wavefunction calculations, recombination probability, transition energy, and optical emission computations quickly and accurately. Contrasting with the existing simulators, QVNTVS is an open-source project and can produce solutions for niche problems like potential wells under an electric field, heterojunctions, recombination, and transition matrices. QVNTVS simulates QWs by solving the Time-Independent Schrödinger Equation for different potential profiles in a discretized space using the finite-difference method and computes the properties of the device using the extracted information from the solution. The results align with the analytical calculations and the experimental data \cite{Gaponenko_Demir_2018}\cite{Savas}.
\end{abstract}


\section{Introduction}
Forming the emissive layer of many semiconductor devices, Quantum Wells (QW) are 2D semiconductor structures employing the confinement phenomenon for their exceptional properties. Inside LEDs, the charge carriers (electrons and holes) rush into the emissive layer QWs to be combined and emit light. Engineering QWs gives the power to alter the operation of semiconductor devices for the need. Analyzing the confinement phenomenon and the behavior of QWs under different setups may give insights about their special properties, and facilitate the research process.
\\
\\ Commercial software and the open-source tools are either not complete for modeling the effects and properties of QWs or are slow and expensive. QVNTVS proposes a solution that is fast compared to the commercial software, accurate as verified by the analytical solutions and experimental data, and is being developed to include other phenomena such as excitons, temperature dependency, and more. Moreover, QVNTVS is an open-source project that can be used by everyone, without license problems.
\\
\\QVNTVS simulates QWs by solving the Time-Independent Schrödinger Equation via a numerical method by discretizing the space. QVNTVS employs the finite-difference method for the solution of the differential equation and uses Ben-Daniel Duke boundary conditions for the heterostructures. Furthermore, QVTVS can handle the QW simulations under biasing conditions imitating actual diode biasing and can output recombination probability and transition energy by calculating overlaps between different wavefunctions and their energy differences. QVTVS will be developed further to include excitons, temperature dependency, and more.
\\
\\In this paper, we will give the theoretical background, methodology, results, and a conclusion on QVNTVS.

\section{Theoretical Background}
\subsection{Solution of Time-Independent Schrödinger Equation inside Quantum Well Heterojunctions}
\label{sec:headings}
Quantum Well structure is a 2D semiconductor structure that creates confinement in one dimension to get special properties raised by confinement, such as discrete energy levels or bands and localized carriers. Quantum well structures are inherently crystal semiconductor lattices. 
\\
\\Since the atoms brought together inside a lattice form continuous energy bands instead of the discrete energy levels in atoms, there are a conduction band, a valence band, and a bandgap in the energy band diagram of the structures. The conduction band is where the charge-carrying electrons live to contribute to the current, and the valence band is where the charge-carrying holes live.
\\
\\Between the conduction and valence bands, there is a bandgap. Inside the bandgap, there can be no states (discarding the indirect recombination). So, switching the bandgap energies between two materials creates potential barriers and wells where only discretized energy levels and wavefunctions live.
\\
\\To start with definitions, the potential wells are low-potential spatial regions sandwiched between high-potential barriers. In QW structures, different materials with different bandgap energies form potential wells. So, there can be one or more potential wells inside the QW structure that can be bent when an electric field is applied. 
\\
\\The importance of the potential wells lies in the confinement phenomenon. The quantum particles governed by the Schrödinger Equation under potentials have bound states, which are so-called confined states, where they become localized. Writing the Time-Independent Schrödinger Equation inside the semiconductor crystals:
\begin{equation}
    -\frac{\hbar}{2m^*}.\frac{d^{2}\psi(x)}{dx^{2}} +V(x)\psi(x) = E\psi(x)
\end{equation} \cite{Miller_2008}
In the first equation, there is an effective mass because the particles obtain effective masses when they are inside the periodic potential structures, such as crystal lattices inside semiconductors. When forming a heterojunction to create a QW, different effective masses should be considered for different materials. Ben-Daniel Duke conditions ensure not only the continuity of the wavefunctions but also the derivative of the wavefunctions divided by the effective masses in this material. Therefore, the heterojunctional structures can be analyzed with the modified Schrödinger equation.
Grouping the wavefunctions at one side, the equation becomes:
\begin{equation}
   ( -\frac{\hbar}{2m^*}.\frac{d^{2}}{dx^{2}} +V(x))\psi(x) = E\psi(x)
\end{equation}
\\The left side is defined as the Hamiltonian operator composed of kinetic energy and potential energy operators, giving the total energy of the system.
\begin{equation}
    \hat{H}\psi(x) = E\psi(x), \hat{H} = \hat{T} + \hat{V} 
\end{equation}\cite{Miller_2008}
\\The equation is an eigen equation now, where the eigenvalues are the energy levels and the eigenfunctions are the wavefunctions.
\\
\\Discretizing the space, the wavefunctions become vectors and the Hamiltonian operator becomes a matrix. Their shapes are determined by the number of intervals that the space is divided into.
\\
\\Since the Schrödinger Equation is a differential equation, a numerical method to solve the differential equation should be employed. QVNTVS uses the finite difference method and computes the derivatives. The first derivative is:
\begin{equation}
    \frac{df(x)}{dx} = \frac{f(x+dx) + f(x)}{dx}, dx:interval\space size
\end{equation}\cite{Wilmott_Howison_Dewynne_1995}
\\The second derivative is:
\begin{equation}
    \frac{d^2f(x)}{dx^2} = \frac{(f(x+dx) - f(x))-(f(x)-f(x-dx))}{(dx)^2}, dx: interval \space size
\end{equation}\cite{Wilmott_Howison_Dewynne_1995}
\\
\\Writing the kinetic energy part of the Hamiltonian operator in this manner and adding the potential function for the indices i (spatial coordinate) in discrete space:
\begin{equation}
    -\frac{\hbar^2}{2}((\frac{\psi_{i+1} - \psi_{i}}{m^*_{i+1/2}dx}) - \frac{(\psi_i-\psi_{i-1})}{m^*_{i-1/2}}) +V_i\psi_{i} = E_i\psi_i
\end{equation}
The effective masses are the means of the 1 over effective masses to correctly model the interfaces where the effective masses change between the materials.
\begin{equation}
    \frac{1}{m^*_{i+1/2}} = \frac{1}{2}(\frac{1}{m^*_i} + \frac{1}{m^*_{i+1}})
\end{equation}
Since the effective masses can change between the interfaces, there shouldn't be any disruptive changes that can harm continuity. Formulating the Hamiltonian
For heterojunctions in discrete space with Ben-Daniel Duke conditions, keep the model eligible.
\\
\\Grouping the i-terms, i+1-terms, and i-1-terms forms the Hamiltonian matrix, where the main diagonal has i-terms, and the off-diagonals have i+1 and i-1 terms.
\\
\\Solving the eigenvalue equation numerically gives the energy values and wavefunctions. Since the radiative recombination needs the bound states, we should consider only the bound states of electrons and holes. So, the recombination probabilities and transition energies can be computed.
\\
\subsection{Recombination Probabilities}

The recombination probabilities between electrons and holes can be calculated by computing the overlap integrals of their wavefunctions.
\begin{equation}
    overlap = \int\psi^*_{electron}(x)\psi_{hole}(x)dx
\end{equation}
And,
\begin{equation}
    Recombination \space  Probability = Overlap^2
\end{equation}
Though we show the analytical counterpart of the overlap function, the computations are done in discrete space.
\\
\subsection{Transition Energies}

To introduce transition energy and optical emission phenomenon, we should define the energy references. For the electrons inside pn-junction semiconductors, we define the electron energy reference to be the lowest point of the conduction band of the well. Similarly, for the holes, we define the hole energy reference to be the topmost point of the valence band. 
\\
\\It should be noted that since the reference point of holes is the topmost point of the valence band of the p-side, the energy levels of holes are negative concerning this point. Because of simplicity, we define hole potentials similarly to the electron potentials. Even though the actual potential is the negated form of it, we can still use the same equation since the effective mass of holes is negative. Let's write negative potentials and masses explicitly:
\begin{equation}
    ( -\frac{\hbar}{-2m^*}.\frac{d^{2}}{dx^{2}} -V(x))\psi(x) = E\psi(x)
\end{equation}
\\Reorganizing the equation:
\begin{equation}
   ( -\frac{\hbar}{2m^*}.\frac{d^{2}}{dx^{2}} +V(x))\psi(x) = -E\psi(x) 
\end{equation}
\\Therefore, we can solve the hole equations as if they were electrons with positive effective mass and potential, then negate the energies.
\\
\\In pn-junctions of diodes, there is a built-in voltage created by the uncompensated charges. The built-in voltage sets a point, after which the diode conducts like a short-circuit. In the QW case, the built-in voltage is created by the energy level differences between the p and n-materials, which are electron transport and hole transport layers in LEDs. Before the applied bias voltage is greater than the built-in voltage, the bands are not straight, and there is an electric field inside. QVNTVS can model both this electric field and the electric field reverse to it for higher biases. 
\\
\\The energy difference between the references can be calculated with:
\begin{equation}
    E_{ref} = E_{bandgap} + qV_{bias} - qV_{built.in}
\end{equation}
\\and
\begin{equation}
    E_{transition} = E_{electron} + E_{ref} -E_{hole}
\end{equation}
\\Therefore, we can compute the energy differences between the respective states of electrons and holes in this manner. Since the recombination probability is already known between all combinations of the wavefunctions, the most probable recombination energies are computed.
\\
\subsection{Optical Emission}
Lastly, we can compute the wavelengths of the transition energies with the photon energy formula.
\begin{equation}
    E_{photon} = \hbar \omega
\end{equation}\cite{Miller_2008}
or
\begin{equation}
    E_{photon} = \frac{hc}{\lambda}
\end{equation}\cite{Miller_2008}
Equating the transition energies that we found in the previous chapter to the wavelength expression, the wavelengths can be computed.
\subsection{Assumptions and Approximations}
To facilitate the design and computations of QVNTVS, we assumed that:
\begin{itemize}
    \item Band gaps are independent of the temperature and the same across the lattice. 
    \item The parabolicity of the potential at the interfaces of heterojunctions can be approximated as a linear shape for low to moderate biases.
    \item Effective masses are constant along the lattice.
    \item The Magnetic effects and spin effects, such as Zeeman, can be discarded.
    \item Recombination probabilities can be computed simply as the spatial overlap integrals.
\end{itemize}
In future versions of QVNTVS, it can be elaborated further to cover the thermal, magnetic, and lattice effects.

\section{Methodology}
\subsection{Initialization}

QVNTVS has a modular and object-oriented approach for the solution written in Python with NumPy \cite{harris2020array}, Matplotlib\cite{Hunter:2007}, and SciPy\cite{2020SciPy-NMeth} libraries. For linear algebra operations, QVNTVS uses NumPy\cite{harris2020array}, for plotting it uses Matplotlib\cite{Hunter:2007}, and for solving the eigen equation it uses SciPy\cite{2020SciPy-NMeth}. Since QVNTVS is an open-source project, other users and developers can add more functions on top of QVNTVS using its modularity. Moreover, the object-oriented approach defines the structures as entities that include the important properties of the materials and the junction in one line of code. Thanks to the object-oriented approach, all operations can be done with only a few lines of code.

\begin{figure}[!htbp]
    \centering
    \includegraphics[width=0.5\linewidth]{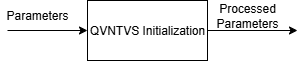}
    \caption{Initalization}
    \label{fig:placeholder}
\end{figure}
Firstly, QVNTVS accepts the parameters needed to simulate the heterojunction quantum wells, such as effective masses, potential well thickness, barrier height, built-in voltage, et cetera. It processes the parameters given at the initialization step and gives out the processed parameters for further steps. The parameter processes include the unit conversions, creating the discretized space, and more.

\subsection{Potential Profiles}
\begin{figure}[!htbp]
    \centering
    \includegraphics[width=0.5\linewidth]{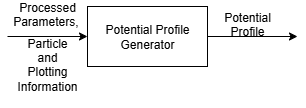}
    \caption{Potential Profiles}
    \label{fig:placeholder}
\end{figure}
QVNTVS accepts the processed parameters from the initialization step to correctly set the potential barriers and wells across the structure. Generating the correct potential holds a key value for accurate simulations. Potential profile generator, which is rectangular and triangular potential profile functions in the code, also accepts the particle information if it is the profile of an electron or hole, and a plotting option.
\\
\\Potential profile generator works by creating a linear space using NumPy\cite{harris2020array}and setting the barriers and wells according to the parameters given at the initialization step. 
\\
\\Triangular well option includes the electric field effects as well. If the quantum wells are biased, the band-bending effects are simulated in the potential profile generating step. The more the electric field and bias, mean more bandbending as a result. 
\\
\\Since the potential profile generator works for both electrons and holes, it is flexible for both structures. As we explained in section 2.3, the hole potentials are created like the electron potentials, though QVNTVS plots them in reverse, because of the reference points. The hole potentials are negative with respect to the reference when there is no net biasing. But, we can solve the Schrödinger Equation while treating holes with positive potentials and masses, then inverting the results. 

Here are some example potential profiles for electrons and holes:
\begin{figure}[!htbp]
    \centering
    \includegraphics[width=0.4\linewidth]{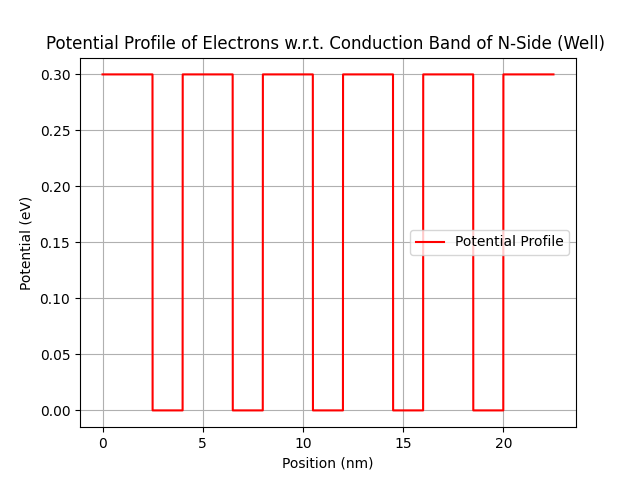}
    \includegraphics[width=0.4\linewidth]{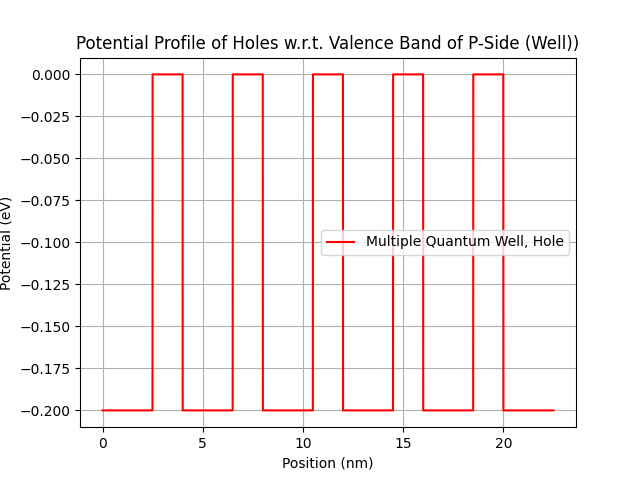}
    \caption{Electron and Hole Potential Profiles}
    \label{fig:placeholder}
\end{figure}
As we see in Figure 3, the electron potentials are positive and the hole potentials are negative with respect to their reference points. Moreover, we can plot tilted and triangular potentials under net biasing, modeling forward or reverse bias modes:
\\
\begin{figure}[!htbp]
    \centering
    \includegraphics[width=0.4\linewidth]{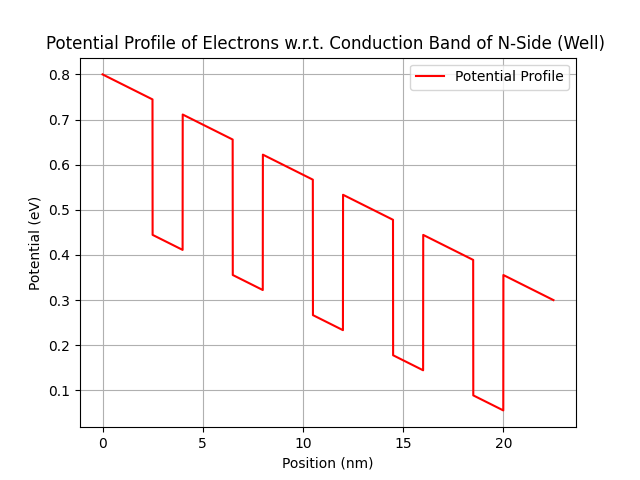}
    \includegraphics[width=0.4\linewidth]{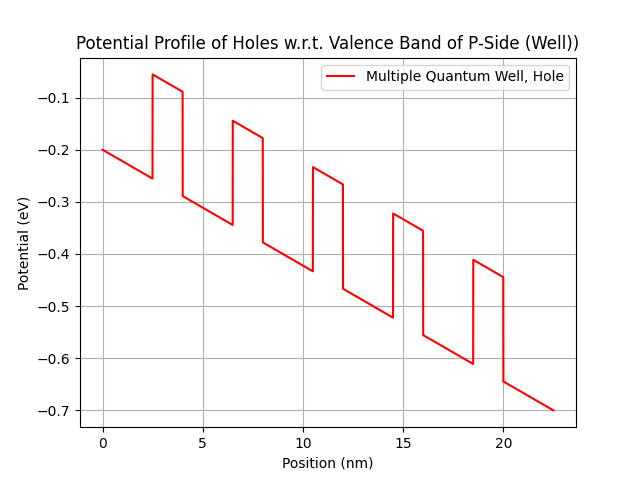}
    \caption{Net Biased Electron and Hole Profiles}
    \label{fig:placeholder}
\end{figure}
\\
Noting that the reference point for the electron potential is the lowest point of the conduction band of the n-side of the well material in pn-junction. The n-side corresponds to the right side of the plot. For the holes, the referencce point is the topmost point of the valence band of the p-side of the well material in pn-junction. The p-side corresponds to the left side of the plot.
\\

\subsection{Effective Masses}
\begin{figure}[!htbp]
    \centering
    \includegraphics[width=0.5\linewidth]{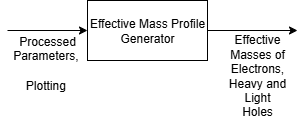}
    \caption{Effective Mass Generator}
    \label{fig:placeholder}
\end{figure}

In the Effective Mass Generator step, QVNTVS accepts processed information of effective masses. It generates effective masses profiles while creating a linear space using NumPy \cite{harris2020array} and defining an efficient mass for every point in space. The effective masses are defined in terms of multipliers of free electron mass. In addition to heavy holes, light hole effective mass profiles can also be generated.
\\
Electrons and holes gain effective masses inside the crystal lattice structures. The effective masses are computed using the second derivative of E(k) function in Brillouin zones. The effective masses can be higher or lighter, positive and negative. Electron effective mass is generally positive since we are concerned with the lowest point of the conduction band and the hole effective mass is generally negative since it is the highest point of the band.

\begin{figure}[!htbp]
    \centering
    \includegraphics[width=0.4\linewidth]{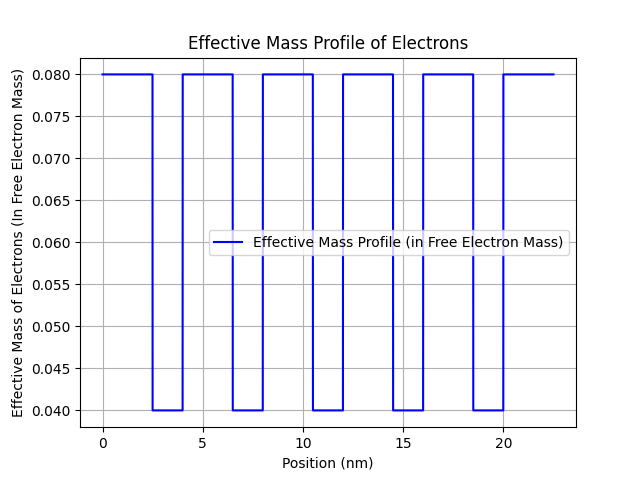}
    \includegraphics[width=0.4\linewidth]{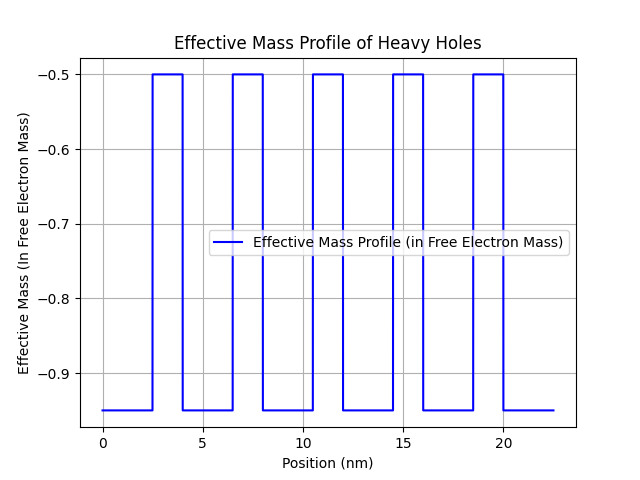}
    \caption{Effective Mass Profiles of Electrons and Holes}
    \label{fig:placeholder}
\end{figure}

As Figure 5 shows, the effective mass profile of electrons is positive and switching between the different materials. Since the barrier-well structure is heterojunctional, it has different materials switching through the structure. These materials have different lattice properties as well, so the effective masses are different. For the holes, we see negative effective masses, though we will compute the energy levels like holes have positive potential profiles and positive effective masses, then negate the overall result.\\
In addition to the effective mass generator, there is also an inverse of mass generator which computes the harmonic mean of the effective masses for each midpoint between the discrete steps in space. It is used for handling the continuity and Ben-Daniel Duke conditions correctly in the finite-difference differential operation step of the Hamiltonian operator.

\subsection{Hamiltonian Matrix}
In the Hamiltonian Matrix Generator step, QVNTVS creates a matrix that corresponds to the matrix in the eigenequation for solving energies and wavefunctions. 
\begin{figure}[!htbp]
    \centering
    \includegraphics[width=0.4\linewidth]{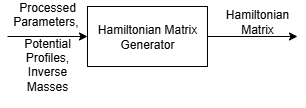}
    \caption{Hamiltonian Matrix Generator}
    \label{fig:placeholder}
\end{figure}
As explained in the section 2.1, Hamiltonian operator defines the total energy of a system, and it is the sum of kinetic energy and potential energy operators. Operators are functions that act on functions in quantum mechanics. \\
To solve the eigenequation of:
\begin{equation}
    \hat H\psi(x) = E\psi(x)
\end{equation}
We should firstly discretize the space to use numerical methods. For better accuracy, the space can be divided into more steps which is controllable in the initialization step.\\
Treating wavefunction as a vector having a value for all discretized steps of the space:
\begin{equation}
    \psi(x) ->\psi_i
\end{equation}
Hamiltonian operator becomes a matrix acting on the wavefunction vector. Writing the eigenequation explicitly for the wavefunction at step i:
\begin{equation}
    -\frac{\hbar^2}{2}((\frac{\psi_{i+1} - \psi_{i}}{m^*_{i+1/2}dx}) - \frac{(\psi_i-\psi_{i-1})}{m^*_{i-1/2}}) +V_i\psi_{i} = E_i\psi_i
\end{equation}
The inverse masses are in the denominators of the wavefunctions. Since the effective masses can change at the interfaces, we took the harmonic mean of the effective masses in section 3.3. As we see in Equation (18), for step-i, there are also i+1 and i-1 terms. So, the Hamiltonian matrix has off-diagonals and main diagonals only. The main diagonal acts on i terms. Grouping i terms together:
\begin{equation}
    \frac{\hbar^2}{2}(\frac{1}{m^*_{i+1/2}}+\frac{1}{m^*_{i-1/2}}) + V_i
\end{equation}
Grouping i+1 and i-1 terms together:
\begin{equation}
    \frac{\hbar^2}{2m^*_{i+1/2}}, \frac{\hbar^2}{2m^*_{i-1/2}}
\end{equation}
Inserting the main diagonals and the off diagonals, we obtain Hamiltonian matrix in discrete space.
\\

\subsection{Solving the Eigenequation}
\begin{figure}[!htbp]
    \centering
    \includegraphics[width=0.4\linewidth]{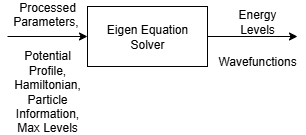}
    \caption{Eigenequation Solver}
    \label{fig:placeholder}
\end{figure}
In this step, QVNTVS solves the eigenequation, which is Equation (16), to get the energy levels and the corresponding wavefunctions.
\\
Eigenequation solver model of QVNTVS takes the processed information from the initialization step, potential profiles, hamiltonian matrix from the previous steps, and particle information and maximum number of levels to solve for. Writing the eigenequation again for discrete space:
\begin{equation}
    \hat H \psi_i = E_n\psi_i
\end{equation}
Using SciPy library in Python, we can solve this eigenequation. The resulting eigenvalues and eigenfunctions correspond to the energy levels and wavefunctions. Though, only the bound levels should be considered as useful for simulation since the main objective is modeling LEDs and LASERs, not the transportation phenomenon. Bound states are defined as the energy levels living inside the wells. For the wells with no slope, the bound states are the energy levels below the potential barrier. For the wells with slope, the bound states are the states with energy levels below the energy barrier between n-side and p-side of the junction. It should be noted that for the bias voltages greater than the built-in voltage, this definition becomes less meaningful, so QVNTVS may not give strictly accurate results for extremely high biases. Nonetheless, the voltage drop on LED can be approximated as the built-in voltage in forward-bias. Therefore, QVNTVS can ensure good approximation.
\\
Here are some example electron energy levels and wavefunctions for n=3 wells with no slope:
\begin{figure}[!htbp]
    \centering
    \includegraphics[width=0.4\linewidth]{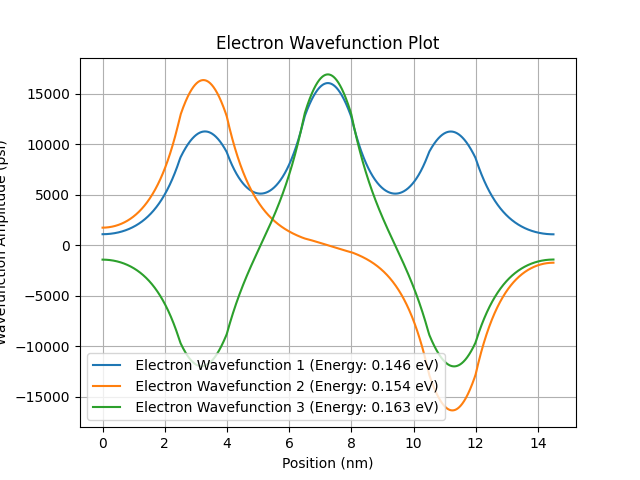}
    \includegraphics[width=0.4\linewidth]{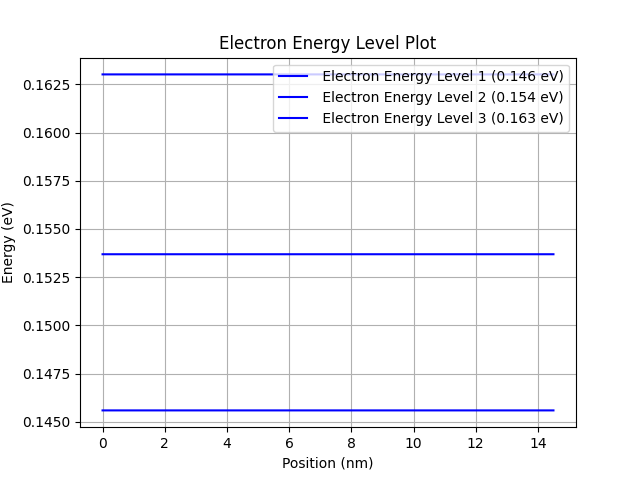}
    \caption{Electron Wavefunction and Energy Levels}
    \label{fig:placeholder}
\end{figure}
\\
It should be noted that the wavefunctions are normalized.
\\
Similarly, here are some example hole energy levels and wavefunctions:
\begin{figure}[!htbp]
    \centering
    \includegraphics[width=0.4\linewidth]{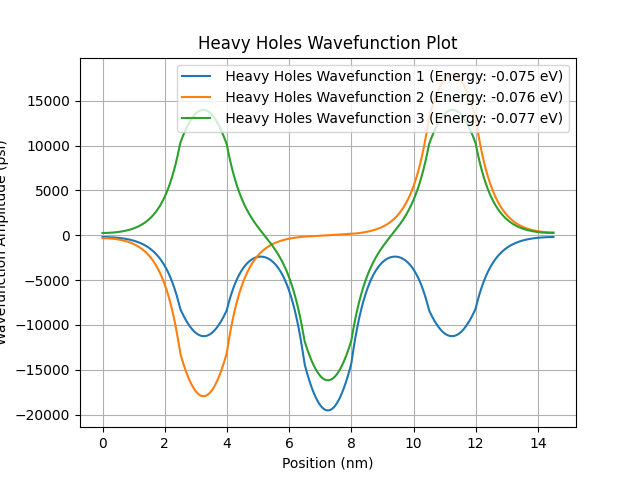}
    \includegraphics[width=0.4\linewidth]{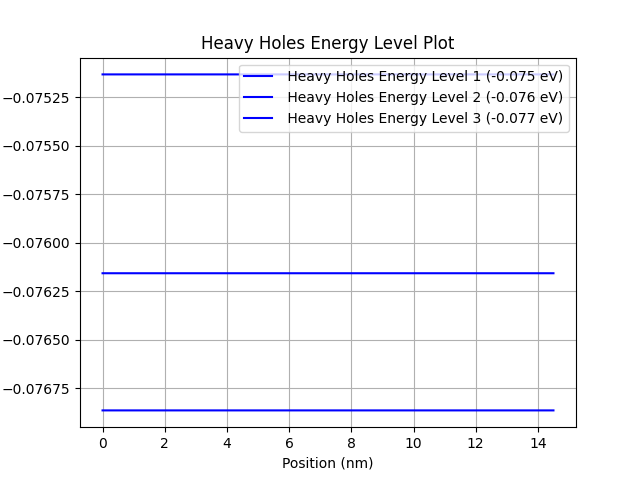}
    \caption{Hole Wavefunction and Energy Levels}
    \label{fig:placeholder}
\end{figure}
\\
The hole energies are negative and increasing to the below. It is natural because of the reference point choice.
\\
For the tilted potentials, energy levels and wavefunctions change because of the bandbending and Stark effect. Here is an example of electron energy levels under net biasing (forward-biasing):\\
\begin{figure}
    \centering
    \includegraphics[width=0.4\linewidth]{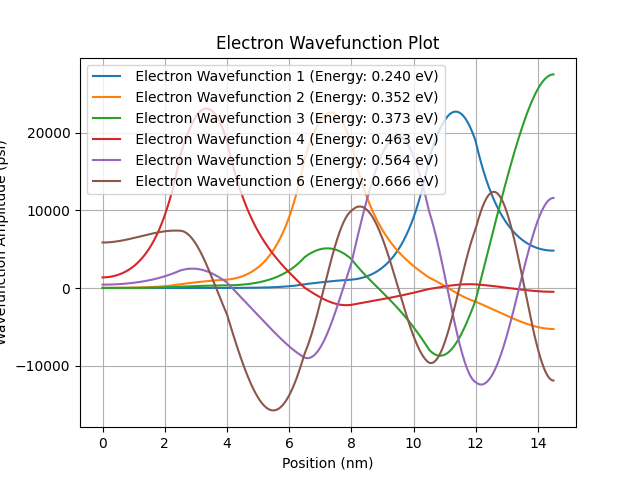}
    \includegraphics[width=0.4\linewidth]{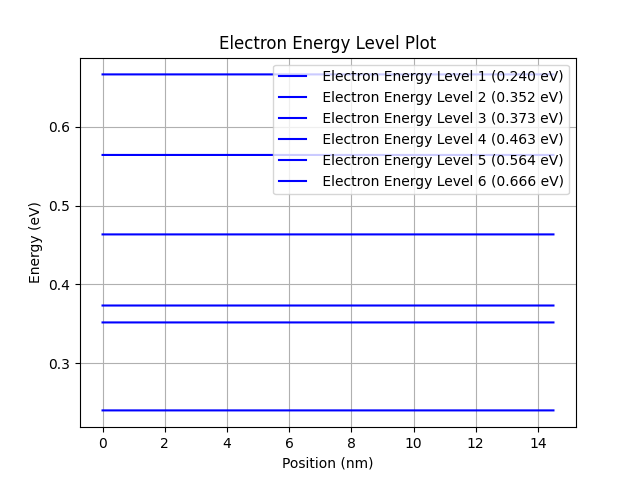}
    \caption{Electron Wavefunction and Energy Levels in Net-Bias}
    \label{fig:placeholder}
\end{figure}
\\As we see in Figure 11, the wavefunctions are not composed of sines and exponentials anymore. They are in Airy functions form which are the analytical solutions of the triangular wells.\cite{Miller_2008} There are more energy levels, because each well is tilted and has different potential than another. So, with respect to the reference, the ground levels have different potential energies.
\\
\\The same idea applies for holes as well. Hole potential is tilted in the same direction with the electron potential, though it has negative values for moderate biases. Therefore, the ground levels of holes are different from each other as well.
\\
\\As we see in Figure 11, also the first ground energy level is changed. This is due to the bandbending and the tilt of the potential. While computing the transition energies, we will take it into account (biasing effect).
\subsection{Recombination Probabilities}
In this step, QVNTVS computes the overlap integral for all combinations of electron and hole wavefunctions.\\
\begin{figure}[!htbp]
    \centering
    \includegraphics[width=0.4\linewidth]{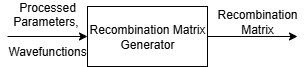}
    \caption{Recombination Matrix Generator}
    \label{fig:placeholder}
\end{figure}
As we see in Figure 12, it takes the processed parameters from the initialization step and the wavefunctions that are eigenfunctions of the Schrödinger Equation. \\
The recombination probabilities should be computed for each combination, because the wavefunctions hold the information about the spatial location of the particle. When there is a strong overlap between the wavefunctions of an electron and a hole, the recombination may occur. It should be noted that, QVNTVS computes the recombination probabilities as the square of the overlap integrals given that there is actually an electron or hole in that state. In reality it is more complex that that, thermal and optical excitation should be taken into account. The next versions of QVNTVS will address that.\\
Here is an example recombination matrix between electron and heavy hole states in Figure 13.
\begin{figure}[!htbp]
    \centering
    \includegraphics[width=0.4\linewidth]{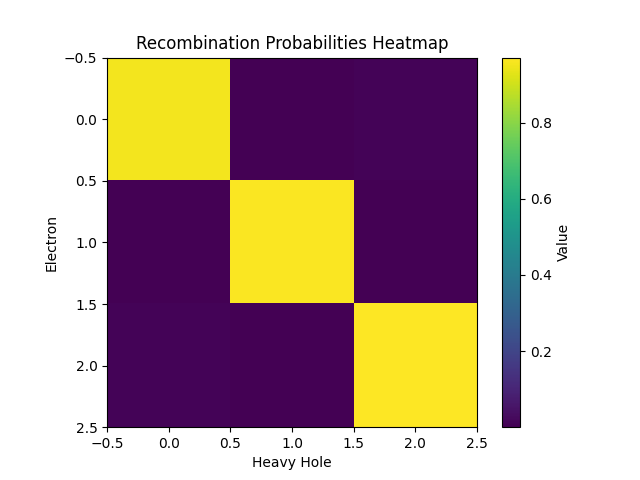}
    \caption{Recombination Heatmap of Electrons and Heavy Holes}
    \label{fig:placeholder}
\end{figure}
It is the recombination matrix of an untilted potential profile. The first states have nearly 100 percent probability of recombination, so the second and the third states given that there are particles in that states.\\
For tilted potentials, the recombination probabilities change:\\
\begin{figure}[!htbp]
    \centering
    \includegraphics[width=0.5\linewidth]{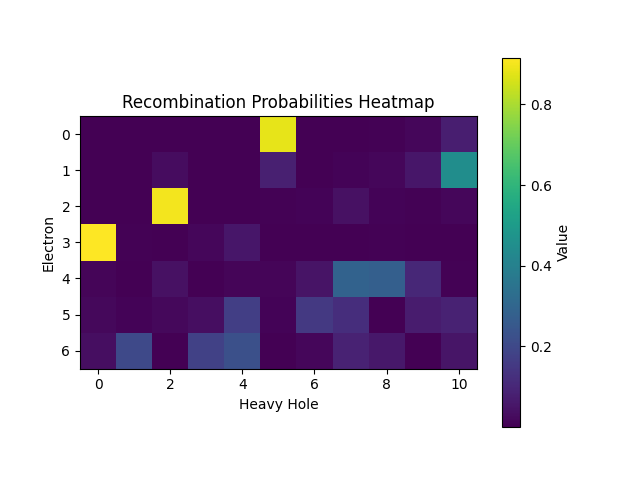}
    \caption{Recombination Heatmap for a Tilted Potential under Net Biasing}
    \label{fig:placeholder}
\end{figure}
\\
In tilted potentials, the carriers should have more energy to cross to the other side. Electrons are injected from n-side and holes are injected from p-side. For their wavefunctions to overlap, their wavefunctions should be extended toward to the other side. Therefore, the recombination probabilities are different.\\
\subsection{Transition Energies}
In this step, the transition energy matrix generator will be investigated.
\begin{figure}[!htbp]
    \centering
    \includegraphics[width=0.4\linewidth]{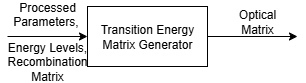}
    \includegraphics[width=0.4\linewidth]{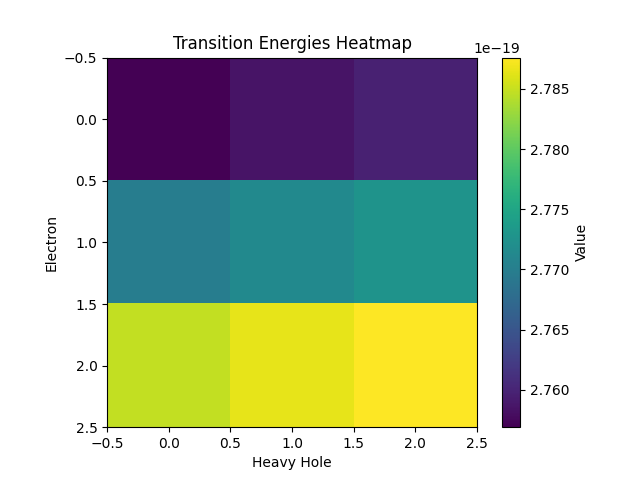}
    \caption{Transition Energy Matrix Generator and its Output}
    \label{fig:placeholder}
\end{figure}
\\
Transition Energy Matrix Generator takes the processed parameters from the initialization step, the energy levels and recombination matrix from the previous sections. \\
QVNTVS matches the wavefunctions with high recombination probability, and calculates the transition energies between them and groups them to create a new matrix having both energy and recombination information. For the potentials with no net biasing, the transition energy is straightforward:\\
\begin{equation}
    E_{transition} = E_{electron} +E_{bandgap}-E_{hole}
\end{equation}
The electrons and holes recombine over the bandgap. So, their transition energies include bandgap as well. In the case of net biasing, calculating the transition energies is not straightforward.\\
For calculating the transition energies, the bandbending should be taken into account. Without any net biasing, the barrier between p-side and n-side of the pn-junction is:\\
\begin{equation}
    E_{barrier} = qV_{built.in}
\end{equation}
\\In the presence of any net biasing, the energy barrier becomes:
\begin{equation}
    E_{barrier} = q(V_{built.in}-V_{biasing}) 
\end{equation}
\\By subtracting barrier energy from the bandgap energy, we get the energy differences between the references of electrons and holes:
\begin{equation}
    E_{ref} = E_{bandgap} - E_{barrier}
\end{equation}
Therefore, the transition energy equation for tilted potentials becomes:\\
\begin{equation}
    E_{transition} = E_{electron} + E_{ref} - E_{hole}
\end{equation}
Though, it should be noted that for extreme biases this model may not give very accurate results. But, the aim of QVNTVS is not simulating semiconductor devices under extreme biases and characterizing them, the aim of QVNTVS is extracting optical, energy levels, and wavefunction information for simulating general device operation.In the next versions of QVNTVS, extreme biasing conditions like Zener breakdown or Avalanche breakdown may be simulated as well.\\
\\For tilted potentials, transition energy matrix becomes:
\begin{figure}[!htbp]
    \centering
    \includegraphics[width=0.4\linewidth]{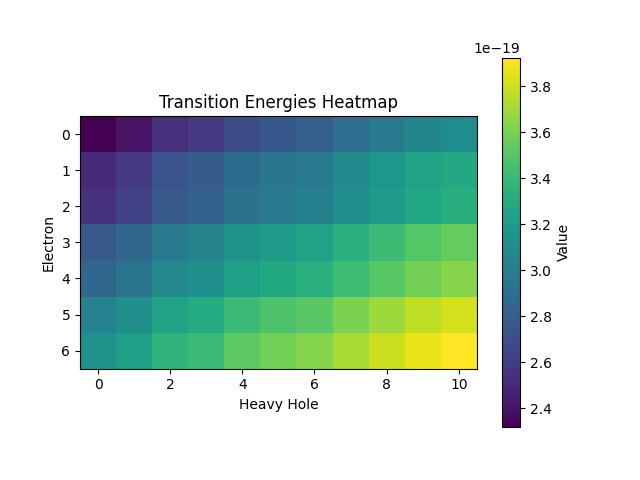}
    \caption{Transition Heatmap for Tilted Potentials}
    \label{fig:placeholder}
\end{figure}
\\As we see in Figure 16, higher energy levels have higher transition energies. For tilted potentials, there is another phenomenon: Generally the transition energies decrease minimally because of the electric field effects. Though, in extreme biases electrons and holes group in the opposite sides of the junction. Even though the recombination probability is decreased, the transition energies are higher, because there is larger energy difference for the particles living at the opposite sides of the junction.

\subsection{Optical Emission Characterization}
\begin{figure}[!htbp]
    \centering
    \includegraphics[width=0.3\linewidth]{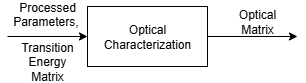}
    \includegraphics[width=0.3\linewidth]{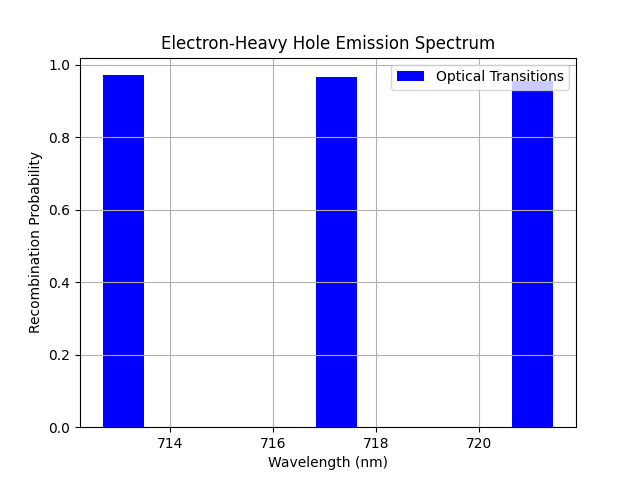}
    \caption{Optical Characterization and its Output}
    \label{fig:placeholder}
\end{figure}
In the optical emission characterization, QVNTVS takes the processed parameters from the initialization step and the transition matrix from the last step including the probabilities.
Since the transition energies are directly related to the emission, we can directly write the photon energy equation:
\begin{equation}
    E_{photon} = \hbar\omega
\end{equation}
or,
\begin{equation}
    E_{photon} = \frac{hc}{\lambda}
\end{equation}
Therefore, we can directly compute the emission wavelength from the transition energies by equating transition energy to the photon energy. It should be noted that, QVNTVS computes the emission wavelengths for the recombination probabilities higher than 30 percent, it gives the dominant emission. Since the transition energies matrix has both energy and probability information inside it, ordering the matrix as a vector by the recombination probabilities gives the transitions with the most probabilities. Then, calculating for energies with probabilities larger than 30 percent directly gives the emission wavelengths. QVNTVS plots the wavelengths by the probabilities to show the dominant wavelengths.\\
\begin{figure}[!htbp]
    \centering
    \includegraphics[width=0.4\linewidth]{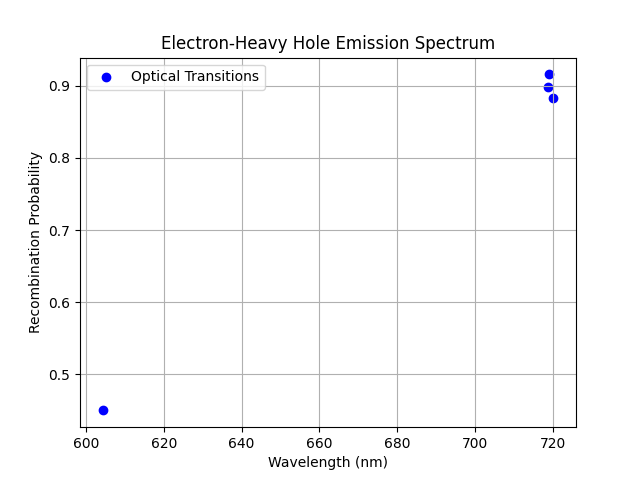}
    \caption{Emission Wavelengths for a Tilted Potential}
    \label{fig:placeholder}
\end{figure}
\\
As Figure 18 shows, there are the wavelengths with the highest probabilities around 720nm as before in the untilted case. Though there is also an emission with a wavelength closer to 605nm with a larger transition energy. This transition is caused by the large electric field caused by the high biasing. Large electric fields drive carriers to the opposite sides at the junction. Since the opposite sides have a greater potential energy difference, the recombination between these states also has greater energy. Though the recombination probabilities between the states at the opposite sides are low because electrons and holes are grouped at the opposite sides.\\
Moreover, QVNTVS assumes that there is an indeed a state in this energy, but the probability of a particle living in this state is governed by thermal and optical excitation. So, the recombination probabilities should taken as the actual values, but the probabilities given that there is a particle in this level. This is the reason for higher emission energies and shorter wavelengths showing up in the simulated spectrum, but they may not be experimentally measured. These phenomena will be addressed in the next versions of QVNTVS as well.

\section{Results}
\subsection{Infinite Rectangular Well}
To verify the results of QVNTVS, we will compare the results with the analytical results. We will run the simulation and solve analytically for 5nm wide wells with 5nm wide barriers, and very high potential barriers imitating the infinite well case. For the easy computation, we take the electron mass to be 1 free electron mass throughout the well and barriers.\\
The equation for calculating the energy levels for the infinite rectangular wells:
\begin{equation}
    E_n = \frac{\hbar^2\pi^2n^2}{2m^*a^2}, a:well.width
\end{equation} \cite{Gaponenko_Demir_2018}

\begin{center}
\begin{tabular}{||c c c||} 
 \hline
 Results & Analytical Solution& QVNTVS \\ [0.5ex] 
 \hline\hline
 First Level & 0.015eV & 0.015eV  \\ 
 \hline
 Second Level & 0.059eV & 0.060eV \\
 \hline
 Third Level & 0.134eV & 0.135eV \\
 \hline 
  & Figure 19: Infinite Well Result Comparison Table & \\[1.0ex]
 \hline
\end{tabular}
\end{center}
As we see in Figure 19, the results for the infinite rectangular wells are within the acceptable error range.\\
By inserting hole masses and negating the formula, hole energy levels can be computed.

\subsection{Finite Rectangular Well Heterojunction}

In this step, we will compare the results to Savas Delikanli, Hilmi Volkan Demir et al. paper \cite{Savas} \\
In the paper \cite{Savas}, CdS/CdSe heterojunction's experimental transition energies are given. We will compare the results for the electron-heavy hole recombinations. We will investigate the transition energies of different well widths, or monolayers (ML).\\

\begin{center}
\begin{tabular}{||c c c||} 
 \hline
 Results: & Experimental Results& QVNTVS \\ [0.5ex] 
 \hline\hline
 2ML-8ML & 2.23eV & 2.27eV  \\ 
 \hline
 3ML-8ML & 2.12eV & 2.17eV \\
 \hline
 4ML-8ML & 2.05eV & 2.09eV \\
 \hline 
  & Figure 20 : Experimental Result Comparison Table & \\[1.0ex]
 \hline
\end{tabular}
\end{center}
As we see in Figure 20, the errors between the experimental results\cite{Savas} are at most 2 percent, which is acceptable.\\
We can argue that QVNTVS outputs are higher than the experimental results. The reason for this can be the approximations and assumptions made to facilitate the simulation process. For more accurate results (error <1 percent), Brillouin zones, thermal phenomenon, and more should be taken into account.\\
Moreover, it should be noted that decreasing the barrier size may increase the error percentage because of the coupling effects. In the next versions of QVNTVS, these phenomena can be incorporated. Nonetheless, we can argue that QVNTVS is an open-source product aiming for fast and accessible simulation, and QVNTVS provides users with results in an acceptable error range.
\section{Conclusion}
To conclude, we introduced QVNTVS with the theoretical background needed, methodology, and results compared to the analytical and experimental solutions. We elaborated the modules, their workflows, and the computations. We showed that the QVNTVS is a successful open-source project for fast, efficient, and accurate simulations, and we have shown that it satisfies the requirements, while having some limitations to be adressed.\\
This project aimed to help researchers and students around the globe, and can be used by everyone who wants to simulate fast, free, and accurately. Moreover, this project is written clearly, and there is a link to the source code in this paper. Therefore, it can be used by students who try to understand quantum mechanics, LEDs and want to see the applications, results of it.\\
For further developments, the author considers adding thermal effects, excitonic phenomena, more complex lattice modeling, and more.
\section{Acknowledgements}
This paper and the related project QVNTVS are the original works by the author, Barbaros Şair. 

\section{Special Thanks}
The author wants to say special thanks to Assoc.Prof.Betül Canımkurbey, Senior Scientist Furkan Işık, and Prof.Hilmi V. Demir from UNAM, National Nanotechnology Research Centre for their support.

\bibliographystyle{ieeetr}  

\end{document}